\documentclass[nofootinbib,amsmath,amssymb,aps,prd,balancelastpage,superscriptaddress]{revtex4-2}

\usepackage{color}
\usepackage[dvipsnames]{xcolor}
\usepackage[active]{srcltx}
\usepackage{amsmath,amsfonts,amssymb,amsthm,amstext,amscd,eucal,srcltx}
\usepackage{epsfig,graphicx,bm}
\usepackage{epstopdf, epsf}
\usepackage{dcolumn}
\usepackage{hyperref}
\usepackage{tensor}
\usepackage{lipsum}
\usepackage{appendix}
\usepackage{tikz}
\usepackage{caption}
\usepackage{subcaption}
\usepackage{color} 

\usepackage{caption}
\usepackage{ragged2e}
\newcommand{\be}{\begin{equation}}
\newcommand{\ee}{\end{equation}}

\newcommand{\bse}{\begin{subequations}}
\newcommand{\ese}{\end{subequations}}
\newcommand{\bea}{\begin{eqnarray}}
\newcommand{\eea}{\end{eqnarray}}
\newcommand{\ba}{\begin{array}}
\newcommand{\ea}{\end{array}}
\newcommand{\bc}{\begin{center}}
\newcommand{\ec}{\end{center}}

\begin{document}
\vspace*{3mm}

\title{Gravitational waves in  minimal Maxwell $f(R)$ gravity}

\author{Salvatore Capozziello}
\affiliation{Dipartimento di Fisica ”E. Pancini”, Università di Napoli “Federico
II”,  Complesso  Universitario  di Monte S. Angelo, Edificio G, Via Cinthia, I-80126,
Napoli, Italy}
\affiliation{Scuola Superiore Meridionale, Largo S. Marcellino 10, I-80138, Napoli,
Italy}
\affiliation{Istituto Nazionale di Fisica Nucleare, Sezione di Napoli, Napoli,
Italy}

\author{Qingyu Gan}
\affiliation{Scuola Superiore Meridionale, Largo S. Marcellino 10, I-80138, Napoli,
Italy}
\affiliation{Istituto Nazionale di Fisica Nucleare, Sezione di Napoli, Napoli,
Italy}

\begin{abstract}
\noindent
We discuss the conversion mechanism from scalar field to gravitational waves in magnetic background in  $f(R)$ gravity minimal coupled to the Maxwell electrodynamics. Applying the conversion in early universe with primordial magnetic field and the neutron star with strong magnetosphere,  the generated gravitational waves, converted from the scalar field, exhibit distinct power spectra which can be potentially probed by future experiments such as DECIGO, BBO, CE and ET.
\end{abstract}

\maketitle

\section{Introduction}

In last decades, a wide range of cosmological observations have revealed the accelerated expansion of the universe, both in its primordial and late stages. These phenomena challenges  General Relativity (GR)  in explaining the cosmos evolution without invoking unknown components such as new scalar particles capable of driving inflation at UV scales and   dark energy at IR scales. The fact that, till now, no new fundamental particle component has been clearly revealed  motivated the exploration of various extended or alternative theories of gravity that, while reproducing the good results of GR, should cure shortcomings at UV and IR scales. See \cite{Faraoni:2010pgm} for a comprehensive discussion. In particular, $f(R)$ gravity is a straightforward extension generalizing  the Einstein-Hilbert action, linear in  the Ricci scalar $R$, to a generic function of it.  This approach preserves the geometric foundations of GR while introducing new degrees of freedom and dynamics capable of addressing many cosmic problems to some extents (see for example  \cite{Capozziello:2002rd, Capozziello:2007ec, DeFelice:2010aj, Nojiri:2010wj, Capozziello:2011et, Sotiriou:2008rp, Nojiri:2017ncd}).

Specifically, it can be demonstrated that $f(R)$ gravity can be recast as  a scalar-tensor theory, equipped with a massive scalar and two massless tensor degrees of freedom  \cite{Capozziello:2008rq,Bogdanos:2009tn,Liang:2017ahj, Capozziello:2018mpw, Katsuragawa:2019uto}. The scalar mode corresponds to the scalar perturbation related to   the form of $f(R)$ function while the tensor modes are the traceless components which constitute, essentially, the gravitational waves  (GWs) of GR.    In vacuum, scalar and tensor fields evolve independently in the linear order perturbations, whereas they are coupled to each other in  higher order regimes. In particular, the first order of scalar perturbation can induce the second order GWs \cite{Zhou:2024doz,Kugarajh:2025rbt,Lopez:2025gfu}. Another possibility to couple the scalar and tensor fields is by the inclusion of the matter. One of the simplest choice to introduce  matter into dynamics  is considering the Maxwell electrodynamics, which is also inspired by the multi-messenger astronomy involving GWs and electromagnetic waves. Incorporating Maxwell electrodynamics into $f(R)$ gravity is explored, for example,  in  Refs. \cite{Mazzitelli:1995mp,Bamba:2008ja,Durrer:2022emo,Papanikolaou:2024cwr}. 

In this paper, we find that  including the Maxwell electrodynamics, even minimally coupled  to $f(R)$ gravity, can  source   GWs and give rise to a clear signature in the power spectrum. As a starting case, we can consider a pure constant magnetic field background with strength $B$. In addition, we can expand the metric perturbation in a  Minkowski background. At  linear order in perturbations, the scalar field $\varphi$ can excite  GWs via the term  $G \varphi B^2$, where $G$ is the Newton constant. In the language of particle physics, such a process can be interpreted as a scalaron (scalar perturbation) propagating through the magnetic field that converts into a graviton (tensor perturbation).  The process is similar to  the graviton-photon and axion-photon conversion in presence of a magnetic background  as reported in Refs. \cite{raffelt1988mixing,Ejlli:2018hke,Addazi:2024kbq,Addazi:2024mii}.

Here we propose the scalaron-graviton conversion in a magnetic field in the framework of a minimal Maxwell $f(R)$ model. However, the probability of conversion  from scalaron to graviton is much suppressed due to the presence of $G$  in the source term $G \varphi B^2$. Thus we need to search for  strong magnetic regimes to enhance the conversion process. There are two interesting potential situations that can be taken into account: the early universe fulfilled with a primordial magnetic field quadratically scaling in redshift \cite{Durrer:2013pga}; neutron stars with strong magnetosphere where the current strongest record reach a $10^{15}$Gs level \cite{kouveliotou1998x}. For the same reason of $G$ suppression, the graviton-photon conversion, in strong magnetic environments, has been explored in Refs. \cite{Fujita:2020rdx,Ito:2023fcr, Hong:2024ofh,McDonald:2024nxj,Dandoy:2024oqg}. 

To estimate the generated GW spectrum, understanding the properties of scalaron is essential. The scalaron mass is constrained in both early and late universe: for example, the  Starobinsky inflation   fixes  $m_a\simeq 3\times 10^{13}$GeV  \cite{Ellis:2015pla}; the  BBN requires $10^{-16}\textrm{eV}<m_a<10^{4}\textrm{eV}$   \cite{Talukdar:2023wjh}, the observed GW speed from  binary systems gives  $m_a<10^{-22}$eV \cite{Lee:2017dox} and so on. 
Some of these  mass ranges seem contradictory to each other and hence demands some mechanism to circumvent it, e.g. the chameleon dressing or a time-varying mass, which can be realized in various $f(R)$ models \cite{Capozziello:2014hia}. 
Since the main purpose of this paper focuses on the scalaron-graviton conversion, we  perform a phenomenological approach taking $m_a$ to be a free parameter in early universe analysis and sufficient small in neutron star analysis. Regarding the density spectrum of the scalaron, we simply regard it as a radiation (at least during the conversion process) constrained by extra effective relativistic species. Assuming  a scale invariant spectrum of scalaron as the simplest scenario, it is possible to show that the generated GWs can be detected by DECIGO \cite{kawamura2011japanese},
BBO \cite{Corbin:2005ny}, CE \cite{Reitze:2019iox} and ET \cite{Hild:2010id}.

The  paper is organized as follows. the first order perturbation  of Maxwell $f(R)$ gravity is given in Sec. \ref{sec:setup}. The scalaron-graviton conversion probability in constant magnetic background is derived in Sec. \ref{sec:conversion}.  Two applications in early universe and neutron star are studied  in Sec. \ref{sec:app}. Discussion and  conclusion are reported in Sec. \ref{sec:con}. 

\section{ Maxwell $f(R)$ gravity}
\label{sec:setup}
Let us consider a generic $f(R)$ model minimally coupled to the Maxwell
electrodynamics with the action given by
\begin{equation}
S  =  \int\sqrt{-g}d^{4}x\left(\frac{1}{\kappa^{2}}f(R)-\frac{1}{4}F_{\mu\nu}F^{\mu\nu}\right),\label{eq:action-Jordan-1}
\end{equation}
where $\kappa=\sqrt{16\pi G}$ and the electrodynamics tensor is $F_{\mu\nu}=\partial_{\mu}A_{\nu}-\partial_{\nu}A_{\mu}$
with the potential vector $A_{\mu}$. The field equations  for metric
$g_{\mu\nu}$ can be obtained from the variation of the action:
\begin{eqnarray}
R_{\mu\nu}f_{R}-\frac{1}{2}g_{\mu\nu}f(R)+g_{\mu\nu}\nabla^{\alpha}\nabla_{\alpha}f_{R}-\nabla_{\mu}\nabla_{\nu}f_{R} & = & \frac{\kappa^{2}}{2}\left(g^{\alpha\beta}F_{\alpha\mu}F_{\beta\nu}-\frac{1}{4}g_{\mu\nu}F_{\alpha\beta}F^{\alpha\beta}\right),\label{eq:fR-Einstein}
\end{eqnarray}
where $f_{R}=df(R)/dR$ and $\nabla_{\mu}$ is the covariant derivative.
The extra scalar mode in $f(R)$ theory can be extracted from the
trace of equation
\begin{equation}
3\nabla^{\mu}\nabla_{\mu}f_{R}+Rf_{R}-2f(R)  =  0\,\label{eq:trace}
\end{equation}
that, with respect to GR, is dynamical.
Note that the RHS of above equation vanishes due to the conformal
property of the Maxwell electrodynamics. With the definition $\Phi=f_{R}$
and $\frac{dV}{d\Phi}=\frac{2f(R)-Rf_{R}}{3}$, Eq. \ref{eq:trace} can
be rewritten in form of the Klein Gordon-type equation
\begin{equation}
\nabla^{\mu}\nabla_{\mu}\Phi-\frac{dV}{d\Phi}  =0  .
\end{equation}

To linearize the theory, we perturb the scalar field as $\Phi=\Phi_{0}+\varphi$
around its ground value $\Phi_{0}$ determined by the minimum of $\left.\frac{dV}{d\Phi}\right|_{\varPhi=\Phi_{0}}=0$.
Moreover, we assume the background spacetime is Minkowski 
and perturb the metric field $g_{\mu\nu}=\eta_{\mu\nu}+h_{\mu\nu}$.
Thus the equation of  $\varphi$  becomes
\begin{equation}
\square\varphi-m_{S}^{2}\varphi  =  0,
\end{equation}
where $\square=\eta^{\eta\nu}\partial_{\mu}\partial_{\nu}$ is d'Alembert operator and its mass is given by $m_{S}^{2}=\left.\frac{d^{2}V}{d\Phi^{2}}\right|_{\varPhi=\Phi_{0}}$.

Regarding the perturbation of  field equations in $f(R)$ theory, it is worth introducing the
variable
\begin{equation}
\overline{h}_{\mu\nu}  =  h_{\mu\nu}-\frac{1}{2}\eta_{\mu\nu}h-\eta_{\mu\nu}\varphi,
\end{equation}
where $h=\eta^{\mu\nu}h_{\mu\nu}$. As demonstrated in Ref. \cite{Capozziello:2008rq,Liang:2017ahj},
$\overline{h}_{\mu\nu}$  describes two massless tensor/polarizations
of GWs as in GR, while $\varphi$ is a massive scalar
field arising from the higher-order of metric $f(R)$ gravity \cite{Capozziello:2011et}. In particle language, perturbations of  $\overline{h}_{\mu\nu}$ and $\varphi$ fields are dubbed ``graviton'' and ``scalaron'', respectively. 
Due to the traceless stress-energy  tensor of the Maxwell electrodynamics,
we are allowed to impose the Lorenz traceless gauge fixing on $\overline{h}_{\mu\nu}$: 
\begin{equation}
    \partial^{\mu}\overline{h}_{\mu\nu}=0, \qquad \eta^{\mu\nu}\overline{h}_{\mu\nu}=0,
    \label{eq:lorenz-gauge}
\end{equation}
and expand the Ricci tensor and LHS of Eq. \ref{eq:fR-Einstein} to
the linear order
\begin{eqnarray}
R_{\mu\nu} & = & \frac{1}{2}\left(-\square \overline{h}{}_{\mu\nu}+2\partial_{\mu}\partial_{\nu}\varphi+\eta_{\mu\nu}\square \varphi\right),\nonumber \\
LHS & = & R_{\mu\nu}\Phi_{0}-\frac{1}{2}g_{\mu\nu}\left(f(0)+f_{R}(0)R+o(R^{2})\right)+\eta_{\mu\nu}\square\varphi-\partial_{\mu}\partial_{\nu}\varphi\nonumber \\
 & = & -\Phi_{0}\frac{1}{2}\eta^{\alpha\beta}\partial_{\alpha}\partial_{\beta}\overline{h}_{\mu\nu}+\left(\Phi_{0}-1\right)\left(\partial_{\mu}\partial_{\nu}\varphi-\eta_{\mu\nu}\square\varphi\right)\nonumber \\
 & = & -\frac{1}{2}\square\overline{h}_{\mu\nu}.
 \label{eq:LHS}
\end{eqnarray}
In the above derivation,  the $f(R)$ model  is required to  satisfy the conditions $\Phi_{0}=f_{R}(0)=1$ and $f(0)=0$. A typical
example is the Starobinsky model $f=R+\alpha R^{2}$. Finally  we obtain the linearized field equations for massless tensor
and massive scalar modes of GWs
\begin{eqnarray}
\square\overline{h}_{\mu\nu} & = & \kappa^{2}\left(\overline{h}^{\alpha\beta}F_{\alpha\mu}F_{\beta\nu}-\frac{1}{2}\eta_{\mu\nu}\overline{h}^{\rho\sigma}F_{\alpha\rho}F_{\;\sigma}^{\alpha}+\frac{1}{4}\overline{h}_{\mu\nu}F_{\alpha\beta}F^{\alpha\beta}-\varphi\left(F_{\alpha\mu}F_{\;\nu}^{\alpha}-\frac{1}{4}\eta_{\mu\nu}F_{\alpha\beta}F^{\alpha\beta}\right)\right), \label{eq:h} \\
\square\varphi & = & m_{S}^{2}\varphi.\label{eq:linear-order}
\end{eqnarray}
Here the index are raised or lowered by $\eta_{\mu\nu}$ and, from now
on, we follow this rule unless  explicitly stated. It is worth mentioning
that the electrodynamics equation is ignored here because we simply
assume a static constant magnetic field. We will expand to include
the dynamics of electromagnetic field (i.e. photons) in a forthcoming paper.

\section{Constant magnetic background}
\label{sec:conversion}
From Eq. \ref{eq:h}, we find that the scalar field interacting
with the electromagnetic one can source the GWs. In fact, the main
purpose of this paper is to show that the scalaron propagating through
a magnetic region can convert into gravitons. In order to demonstrate this statement, let us 
 consider  scalaron and graviton propagating in a Minkowski spacetime
along the $Z$-direction, denoted as $\varphi(t,Z)$ and $\overline{h}_{\mu\nu}(t,Z)$, in presence of a static constant magnetic background.
The direction of the magnetic field is assumed to be fixed in the
transverse $X$-$Y$plane, for instance along $Y$-direction, namely
$\boldsymbol{B}=(0,B,0)$. In the transverse traceless gauge with $\overline{h}_{0\mu}=0$, $\partial_{i}\overline{h}^{i}_{\,j}=0$ and  $\overline{h}^{i}_{\,i}=0$ 
\footnote{Rigorously, transverse traceless (TT) gauge of $\overline{h}_{\mu\nu}$
can always be imposed in vacuum but not in general cases with non-vanishing
sources. The traceless stress-energy  tensor of the Maxwell electrodynamics allows a gauge fixing at most to Eq. \ref{eq:lorenz-gauge}. Using the TT gauge can
be regarded as a projection onto the tensorial components of $\overline{h}_{\mu\nu}$
whereas the   others are ignored. See Refs. \cite{Ejlli:2018hke,Ejlli:2020fpt,Hwang:2023nqx,Fujita:2020rdx}
for comments on the TT gauge in relevant topics.}
, the graviton  can
be decomposed into plus and cross polarizations as $\overline{h}_{ij}=\overline{h}_{+}e_{ij}^{+}+\overline{h}_{\times}e_{ij}^{\times}$,
where bases are $e_{ij}^{+}=X_{i}X_{j}-Y_{i}Y_{j}$ and $e_{ij}^{\times}=X_{i}Y_{j}+X_{j}Y_{i}$.
Projecting $e_{ij}^{+}$ and $e_{ij}^{\times}$ on Eqs. \ref{eq:h}-\ref{eq:linear-order}
and using $F_{0\nu}=0,F_{ij}=\varepsilon^{ijk}B_{k}$, we obtain
\begin{eqnarray}
\left(-\partial_{t}^{2}+\partial_{Z}^{2}\right)\overline{h}_{+}(t,Z)-\frac{1}{2}\kappa^{2}B^{2}\overline{h}_{+}(t,Z) & = & -\frac{1}{2}\kappa^{2}B^{2}\varphi(t,Z),\label{eq:hplus}\\
\left(-\partial_{t}^{2}+\partial_{Z}^{2}\right)\overline{h}_{\times}(t,Z)-\frac{1}{2}\kappa^{2}B^{2}\overline{h}_{\times}(t,Z) & = & 0,\label{eq:hcross}\\
\left(-\partial_{t}^{2}+\partial_{Z}^{2}\right)\varphi(t,Z)-m_{S}^{2}\varphi(t,Z) & = & 0.\label{eq:scalar}
\end{eqnarray}
The presence/absence of the source term on the RHS of the first two equations
is caused by the particular choice of the magnetic field direction,
leading to different dynamics of the two polarizations. Since the magnetic
effect on polarization is out of the scope of this paper, thus we
simply ignore the $\overline{h}_{\times}$ term and denote $\overline{h}_{+}$
as $\overline{h}$ (not the trace of $\overline{h}_{\mu\nu}$). 

The term $\kappa^{2}B^{2}\overline{h}$ gives rise to an effective
mass of the graviton, $m_{GW}=\kappa B/ \sqrt{2}$. We work in the
relativistic regime where the frequency of particles is much larger
than their mass, namely $\omega\gg m_{GW}$ and $\omega\gg m_{S}$.
Performing the Fourier transformation
\begin{eqnarray}
\overline{h}(t,Z) & = & \int d\omega\widetilde{h}(\omega,Z)e^{-i\omega t},\nonumber \\
\varphi(t,Z) & = & \int d\omega\widetilde{\varphi}(\omega,Z)e^{-i\omega t},
\end{eqnarray}
and applying the Slowly Varying Envelop Approximation \cite{Ejlli:2018hke} (valid for $\omega \gg \kappa B/2 $), $\left(-\partial_{t}^{2}+\partial_{Z}^{2}\right)=\left(\omega^{2}+\partial_{Z}^{2}\right)=2\omega\left(\omega+i\partial_{Z}\right)$,
we reduce Eqs. \ref{eq:hplus} and \ref{eq:scalar} to the linear
differential equations 
\begin{eqnarray}
i\partial_{Z}\widetilde{h}(\omega,Z)+\left(\omega-\frac{m_{GW}^{2}}{2\omega}\right)\widetilde{h}(\omega,Z) & = & -\frac{\kappa^{2}B^{2}}{4\omega}\widetilde{\varphi}(\omega,Z),\\
i\partial_{Z}\widetilde{\varphi}(\omega,Z)+\left(\omega-\frac{m_{S}^{2}}{2\omega}\right)\widetilde{\varphi}(\omega,Z) & = & 0.
\end{eqnarray}
For the boundary condition $\widetilde{h}(\omega,Z_{i})=0$, the solution
is 
\begin{equation}
\widetilde{h}(\omega,Z)  =  e^{i\omega d}\left(1-e^{-i\frac{\kappa^{2}B^{2}}{4\omega}d}\right)\widetilde{\varphi}(\omega,Z_{i}),
\end{equation}
where $d=Z-Z_{i}$ is the traveling distance. The conversion rate
for scalarons converting into gravitons at a distance $d$ is
\begin{equation}
\mathcal{P}_{S\rightarrow GW}(d)  = \frac{\left|\widetilde{h}(\omega,Z)\right|^{2}}{\left|\widetilde{\varphi}(\omega,Z_{i})\right|^{2}}=4\textrm{sin}^{2}\left(\frac{\kappa^{2}B^{2}}{8\omega}d\right).
\label {eq:P}
\end{equation}
Assuming the initial energy density of the scalar source as $\Omega_{\varphi}(\omega,Z_{i})$,
we can estimate the abundance of GWs converted from the scalarons in presence
of a magnetic field as
\begin{equation}
\Omega_{GW}(\omega,Z)  = \Omega_{\varphi}(\omega,Z_{i})\mathcal{\mathcal{P}}_{S\rightarrow GW}.
\end{equation}
We stress that $\mathcal{P}$ is properly called ``conversion rate''
rather than ``conversion probability'', because such a conversion
process is unidirectional only from scalarons to gravitons. Thus it can be larger than unity, reaching to maximum 4 from the above expression. Nevertheless, in
most physical circumstances, $\mathcal{P}$ is much smaller than the
order of unity. Such a suppressed conversion rate leads to
a negligible effect on the scalar source, and hence allows one to use always
the Klein Gordon  equation to describe the scalaron as a free field. Therefore,   we simply interpret $\mathcal{P}$ as the conversion probability  converting from scalarons to gravitons when $\mathcal{P}\ll 1$, and express Eq. \ref{eq:P} as   
\begin{equation}
\mathcal{P}_{S\rightarrow GW}  =  \frac{\kappa^{4}B^{4} d^2}{16\omega^2}.
\label {eq:Psmall}
\end{equation}

The conversion mechanism from scalarons to gravitons proposed in this paper is reminiscent of other mixing systems widely studied in literature, for instance, the photon-graviton mixing in magnetic field. Ignoring the plasma effect on photons, the probability of photon-graviton conversion is $\mathcal{P}=\textrm{sin}^2 (\kappa Bd)$, which reduces to $\mathcal{P}=\kappa^2B^2d^2$ when $\mathcal{P}\ll 1$ \cite{Addazi:2024kbq}. Comparing with Eq. \ref{eq:Psmall}, the scalaron-graviton conversion has a quartic  dependence on the magnetic field rather than the conventional quadratic standard of the photon-graviton case. Moreover, the extra factor $\kappa ^2 B^2/16 \omega^2<1$ indicates a suppressed  probability of scalaron-graviton conversion than the photon-graviton case. 
 
\section{Applications}
\label{sec:app}
Let us study now the possible  applications of the scalaron-graviton conversion in the contexts of cosmology and astrophysics. The quartic power of magnetic strength in Eq. \ref{eq:Psmall} 
implies a suppressed conversion probability in the weak magnetic regimes. For instance, considering the scalaron propagating  at a distance $\sim \textrm{Mpc}$ in the intergalactic medium with strength $\sim \textrm{nGs}$, the conversion probability is estimated  to be of the order $\mathcal{P}\sim 10^{-60} (\textrm{Hz}/\omega)^2$, which is extremely small in a reasonable frequency region. To have a significant amplification of conversion probability, we need to focus on  regimes  with strong magnetic fields. Two possibilities are explored in the following: the early universe and the magnetar.

\begin{figure}
    \centering
    \includegraphics[scale=0.7]{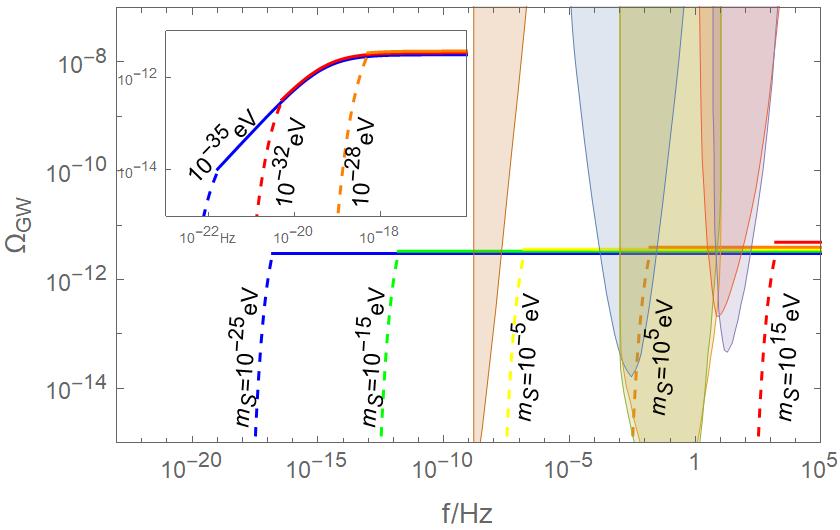}\includegraphics[scale=0.7]{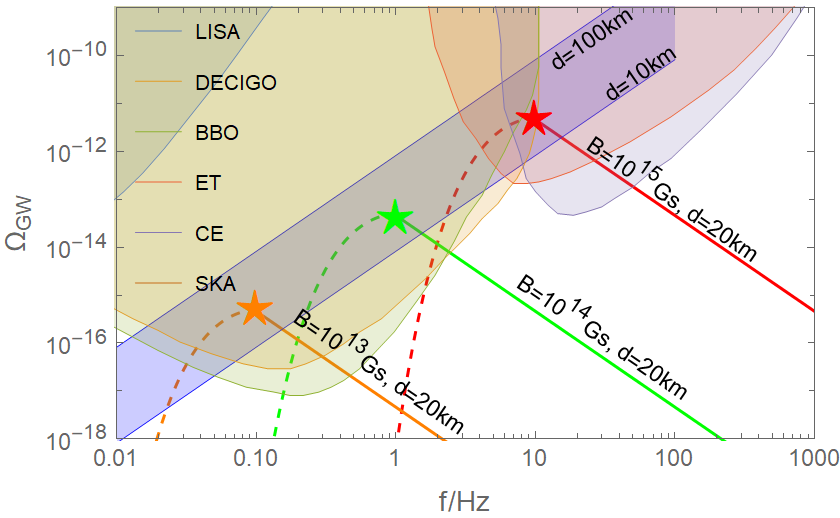}
    \caption{
    The power spectrum of GWs converted from the scalaron propagating in the primordial magnetic field in  early universe (Left) and the strong magnetic field generated by  magnetars (Right). The scalaron source is assumed to be in a scale invariant spectrum with density $\Omega_\varphi=0.06 \Omega_\gamma$ in the whole post-inflationary epoch. The converting GW spectra are compared to various experiments with their sensitivity curves obtained from \cite{Schmitz:2020syl}.  The blue shade in right panel denotes the maximum of GW density by varying magnetic strength of the magnetar and effective magnetosphere size. The damping effect is expected to be dominant when $f\lesssim \sqrt{\kappa B_0 m_S}/2 \pi $ (Left) or $f\lesssim \kappa B /2\pi$ (Right), which is  schematically shown in dashed lines.  }
    \label{fig:SGW}
\end{figure}

In most literature,  various $f(R)$ models are constructed to explain the inflation or the dark energy, meanwhile the model parameters are constrained by the observational data. As mentioned in Introduction, our main purpose is to take into account  the conversion mechanism from  scalarons to gravitons in a magnetic background, thus we adopt a phenomenological approach to study its potential effects. We assume the existence of scalarons arisen from $f(R)$ gravity models compatible with  Eq. \ref{eq:LHS} and magnetic field in the whole post-inflation era. Moreover, the scalaron mass $m_S$ is taken to be a free parameter. The magnetic field is expected to be scaled in high redshift $z$ as $B=B_0z^2$, where   $B_0= \textrm{nGs}$  is the strength of the intergalactic  magnetic field today \cite{Durrer:2013pga}. Especially, it is well within the bound from the BBN constrains $B_{\textrm{BBN}}<2\times 10^{11}\textrm{Gs}$ \cite{kawasaki2012updated}. Within one Hubble time, the traveling distance is $d=H^{-1}$ and hence conversion probability is
\begin{equation}
    \mathcal{P}_{S\rightarrow GW}(z)=\frac{1}{16} \left(\frac{\kappa B}{\omega}\right)^2 \left( \frac{\kappa B_0}{H_0} \right)^2, \,\,\,\,\,\,\,\, \omega \gg \kappa B /\sqrt{2},
    \label{eq:PearlyU}
\end{equation}
where the frequency scales as $\omega=\omega_0z$ and Hubble parameter $H=H_0z^2$ in R.D. era. In addition, the above formula is only valid at $\omega \gg m_S$. With these conditions, the conversion probability reaches a maximum $\mathcal{P}\simeq  6\times 10^{-11}$ at $\omega \simeq  \kappa B$. To estimate the generated GW energy density, we consider the simplest case with a scale invariant spectrum of the scalar source. The scalaron can contribute as  a radiation component, constrained by  extra effective relativistic species $\Delta N_{\textrm{eff}}$
with $\Omega_\varphi /(0.2\Omega_\gamma) <\Delta N_{\textrm{eff}}$ \footnote{For a sufficient large mass, scalaron becomes cold DM below a certain temperature in the early universe. Its density abundance can be larger than the radiation one, but there may exist other constrains from the cold DM sector. The different mass cases in left panel of Fig. \ref{fig:SGW} serves more as a reference.}. The Planck constraint
	$\Delta N_{\textrm{eff}} <0.3$ gives rise to the bound $\Omega_\varphi(f)\simeq\Omega_\varphi<0.06 \Omega_\gamma $ \cite{Planck:2015zrl} from BBN. Since scalaron is assumed to be a dark radiation field, thus the ratio to SM radiation is fixed during the evolution of universe. In the RD epoch, $\Omega_\gamma$ approaches to unity and decreases to $\Omega_\gamma \simeq 5\times 10^{-5}$ at present day. We simply assume $\Omega_\varphi(f)$ at the constraint bound $\Omega_\varphi(z)=0.06 \Omega_\gamma(z)$ in the whole post-inflationary epoch, and show several cases of different scalaron masses in the left panel of Fig. \ref{fig:SGW}.  As a result, the generated GWs can be  present in an almost truncated scale invariant power spectrum with energy density $\Omega_{GW}(f)\simeq 10^{-12}$, which could be probed by various detector such as SKA, LISA, DECIGO and BBO. In the case with mass $m_S\lesssim 10^{-28}$eV, the maximum conversion, which takes place at $z\simeq \omega_0/\kappa B_0$, occurs after the RD-MD equality. As a consequence,  $\Omega_\varphi(z)$ is suppressed, and and the spectrum exhibits a $\omega^4$ scaling decay at low frequencies.
In addition, the condition $ \omega \gg m_S $ gives raise to an IR truncate of the spectrum, constraining the frequency region of the spectrum to $f \gtrsim  \sqrt{\kappa B_0 m_S}/2 \pi$. For lower frequency regions, Eq. \ref{eq:P} fails and one needs to solve full equation set Eqs. \ref{eq:hplus} and \ref{eq:scalar}. Nevertheless, we anticipate a damping effect for $\omega<m_S$  because  it  might correspond to the evanescent mode of a localized  field in potentials or interactions.

The other potential possibility to enhance the conversion process happens in the magnetosphere of the  magnetar, a type of neutron star with very strong magnetic field, where the magnetic field can reach up to $B=10^{13-15}$Gs \cite{Hong:2024ofh}. In addition, the  size of magnetosphere, $\sim 10^5 (P/\textrm{s} ) \textrm{km}$ with spin periods $P$, is generally much larger than the star radius ($\sim 10$km) and the magnetic strength decays with distance as $\sim r^{-3}$ \cite{Ito:2023fcr,McDonald:2024nxj}.  To avoid the complexity and perform an order-of-magnitude estimation, we effectively model scalarons passing through the magnetosphere  as a path with constant magnetic field $B$ and  length $d$. From Eq. \ref{eq:Psmall},  the conversion probability, presenting in a power law scaling $\omega^{-2}$, indicates a maximum at $\omega \simeq \kappa B$ in the region $\omega \gtrsim \kappa B$. 
Again, we consider a scale invariant spectrum of the scalaron  $\Omega_\varphi(f)=0.3\times10^{-5}$ but its mass is smaller than the frequency of interests $m_S < 10^{-17}$eV.   In a reasonable range of $B<10^{16}\textrm{Gs}$ and $10\textrm{km}<d<100\textrm{km}$, the maximum energy density of  GWs converted from scalarons locates inside the blue shade region in the right panel of Fig. \ref{fig:SGW}. 
The profiles of the GW spectrum are shown in three benchmarks with $B=10^{13,14,15}$Gs and $d=20$km. The generated GWs can possess energy density in the range $\Omega_{GW}\sim 10^{-16}-10^{-12}$ around frequency region $f\sim 0.1-10$Hz, which can be probed by the future experiments, for instance DECIGO, BBO, ET and CE.
We expect a possible damping feature in the low frequency region $\omega<\kappa B$ (dashed lines) caused by the GW scattering in the strong magnetic medium, while this statement  needs further detailed analysis.  

Two comments are in order. Firstly, as addressed in Ref. \cite{Bamba:2018cup},  it is justified to consider GW propagating through magnetic medium in the flat spacetime background under two conditions $\kappa B<\omega $ and $\kappa B L<1$, where $L$ is the size of the magnetic region. The latter means that the background curvature, sourced by  the magnetic field, can be ignored. These conditions are generally satisfied in our model applications in early universe and  magnetar. Secondly, the above analysis is based on a scale invariant spectrum of scalaron source and can be easily extended to other forms of spectrum. For instance, one can consider a spectrum  peaked at $\omega_p$  with a narrow frequency band $\Delta \omega$: the scalaron density is allowed to increase to  $\sim \Omega_\gamma \omega/\Delta \omega$ in the peaked band. Accordingly,  the abundance of generated GWs can be enhanced by a factor $\sim \omega/\Delta \omega $ in the peaked band, rendering its detectable well within the experimental sensitivities.

Previous discussion  focused on a single conversion channel  from scalaron to GW, and now we consider more general physical scenario involving multiple conversion channels among scalaron, graviton and photon.  Assuming the ingoing GW and photon with initial energy density  $\Omega_{GW}(f,Z_i)$ and $\Omega_{\gamma}(f,Z_i)$ pass through the magnetized region from $Z_i$ to $Z$, the outgoing GW energy density is  
\begin{equation}
	\Omega_{GW}(f,Z)  = \left( 1- \mathcal{P}_{GW\rightarrow \gamma}  \right)\Omega_{GW}(f,Z_i) + \mathcal{P}_{\gamma \rightarrow GW}  \Omega_{\gamma}(f,Z_i)+ \mathcal{P}_{S\rightarrow GW} \Omega_{\varphi}(f,Z_{i}).
	\label{eq:GWnonzeroinital}
\end{equation}
Note that there is no term $\mathcal{P}_{GW \rightarrow S}$ because the scalaron-GW conversion is  unidirectional. The conversion probability between GW and photon is strongly suppressed within frequency region of our interests in this work (i.e. $f\lesssim\mathrm{kHz}$). To see this,  we first estimate the conversion probability as
\begin{equation}
	\mathcal{P}_{GW \leftrightarrow \gamma}\simeq \kappa ^2 B^2 l_{osc}^2 \sin ^2 (d/l_{osc}) \simeq \kappa ^2 B^2 f^2 \omega_{pl}^{-4}, 
\end{equation}
where  $\omega_{pl}$ is the plasma frequency. The last equality holds  because in both early universe and magnetar cases, the size of interested region $d=Z-Z_i$ is much larger than the oscillation length $l_{osc}\simeq f/\omega_{pl}^2$ for $f\lesssim\mathrm{kHz}$ \footnote{In fact, in the physical regimes considered here, the characteristic frequency of particle mixing   is smaller than the plasma frequency of magnetic medium. Consequently,  the photon, unlike graviton and scalaron,  can not propagate freely within the medium, which effectively suppresses the GW-photon conversion. This is also one of the reason that we neglect the electrodynamics from the start of this work. Nevertheless, for completeness, we provide a rough estimate of how small this effect would be if the conversion were to occur. In the magnetoshpere of magnetar, the plasma frequency is $\omega_{pl}\simeq (B/10^{12}\mathrm{Gs})^{1/2} (1 \mathrm{s}/P)^{1/2} \mathrm{GHz}$, where $P$ denotes the spin period \cite{Hong:2024ofh}. For   $B\simeq 10^{12}\mathrm{Gs}$ and $f\lesssim \mathrm{kHz}$, this yields  $l_{osc}\lesssim 10^{-5}\mathrm{m}$, which is much less than the typical magnetosphere size ($\sim \mathrm{km}$). In the early universe with redshift $z$, the physical  plasma frequency is $\omega_{pl} =\sqrt{e^2 n_e/m_e}  \simeq z^{3/2} \mathrm{Hz}$ with electron density $n_e\simeq 0.2 z^3 \mathrm{m}^{-3} $. The corresponding   physical oscillation length is $l_{osc}\simeq  10^{-7}z^{-2}(f/\mathrm{Hz})\mathrm{pc}$, which is much less than the Hubble radius $d \simeq H_0^{-1} z^{-2}$ \cite{Dolgov:2012be}.   }. Thus even for an extreme magnetar with   $B\simeq 10^{15}\mathrm{Gs}$ and $d\simeq 20\mathrm{km}$, the conversion probability remains highly  suppressed $\mathcal{P}_{GW \leftrightarrow \gamma} \lesssim 10^{-20} $ in the low frequency region  $f\lesssim \mathrm{kHz}$. Similarly,  the conversion probability in the early universe is also as tiny as $\mathcal{P}_{GW \leftrightarrow \gamma} \lesssim 10^{-38} $ for  extragalactic  magnetic field $B\simeq \mathrm{nGs}$  and frequency $f\lesssim\mathrm{kHz}$ (in comoving framework).  Therefore, throughout this work,  we can safely neglect the GW-photon conversion terms  in Eq. \ref{eq:GWnonzeroinital}, and conclude that the outgoing GW spectrum is a simple superposition of the ingoing GW component and the GW generated from the scalaron, namely
\begin{equation}
	\Omega_{GW}(f,Z)  = \Omega_{GW}(f,Z_i) + \mathcal{P}_{S\rightarrow GW} \Omega_{\varphi}(f,Z_{i}).
	\label{eq:GWnonzeroinitalsimp}
\end{equation}
 We can apply this conclusion to the case of a stochastic GW background, originating either from  inflationary process or cosmological phase transition \cite {Caprini:2018mtu, Oikonomou:2024geq, Oikonomou:2024jms, Addazi:2023jvg, Oikonomou:2025jmy}. As such a stochastic background propagates through the early universe and(or not) the magnetar in late universe, the resulting GW spectrum  is simply the sum of the original stochastic component and the scalaron-induced GW component as shown in Fig.  \ref{fig:SGW} (the left panel used to add scalaron-induced GW from the early universe to the original GW background; the right panel used to add scalaron-induced GW from the magnetar to the original GW background).  
It is worth to mention that within the framework of Einstein GR, Eq. \ref{eq:GWnonzeroinitalsimp} reduces to $\Omega_{GW}(f,Z)  = \Omega_{GW}(f,Z_i)$, indicating that stochastic GWs with $f\lesssim \mathrm{kHz}$ remain essentially unaltered when traversing strong magnetic environments, whether in the early universe or in  magnetars in galaxies.

Regarding the anisotropy, again for the GR case, the arguments above suggest a strongly suppressed  effect for GW-photon conversion impacting  on the anisotropy of the stochastic GW background within $f\lesssim \mathrm{kHz}$ even when  GWs propagate through magnetars  in galaxies. In contrast, for  $f(R)$ gravity, as shown in the right panel in Fig. \ref{fig:SGW}, magnetars have a significant effect on the  anisotropy of the GW spectrum only in frequency region $0.01\mathrm{Hz} \lesssim f \lesssim 100\mathrm{Hz}$. The  detailed analysis on the anisotropy power spectrum is challenging and is left for future study. Nevertheless, we can roughly estimate on the main angular scale of the anisotropy based on the distribution of magnetars. Firstly,  most known magnetars are observed within our Galaxy, implying the anisotropy spectrum would receive a particular contribute from the Galactic plane \cite{Kaspi:2017fwg}.  For extragalactic magnetars, the anisotropy is expected to be more homogeneous. Observations indicate that around half of the stellar mass observed in galaxies today are formed in just $\sim 3.5\mathrm{Gyr}$ in $z\sim 1-3$ \cite{forster2020star}. Assuming that magnetar formation closely traces the formation history of massive star and supernovae, their abundance is expected to peak during the same epoch. On the other hand,  the magnetically active phase of a magnetar lasts only  $\sim 10^4\mathrm{yr}$ \cite{Kaspi:2017fwg}, thus the number of active magnetars in each galaxy is expected to be largest around $z\sim 1-3$,  which would enhance the induced anisotropy in the stochastic GW background.  For concreteness, we adopt a fiducial redshift $z\simeq 2$. In addition, the mean separation between nearby galaxies is of order $\sim \mathrm{Mpc}$ ($\sim10-100$ times the typical galaxy size), corresponding to an angular separation of $\sim 0.01^\circ$. Therefore,  the angular power spectrum of anisotropies in the   stochastic GW background caused by the scalaron-GW conversion in magnetars is expected to  peak at $\sim 0.01^\circ$.

\section{Conclusion}
\label{sec:con}
In this paper we considered  $f(R)$  gravity minimal coupled to the Maxwell electrodynamics, where the system contains a massive scalar field (scalaron), a massless tensor field (graviton/GW) and an electromagnetic field (photon). As a first approximation, we assumed the electromagnetic field to be a  constant magnetic field as a background, and focused on the dynamics of the scalaron and the graviton in the Minkowski spacetime. We demonstrated that scalarons propagating in a magnetic field can convert into gravitons. This conversion mechanism is analogous to the graviton-photon, axion-photon conversion widely studied in literature. However, there are two distinct features: the conversion is unidirectional, thus proper to call it ``scalaron-induced GW"; the conversion probability has a quartic dependence on the  magnetic field. We perform a phenomenological approach to apply this mechanism in the early universe with a primordial magnetic field and the   neutron star with a strong magnetosphere. In both cases, the generated GWs in a broad range of parameters can be probed by future experiments (see Fig. \ref{fig:SGW}).

Several interesting aspects can be explored in  future researches. An immediate generalization is to include the dynamical   electromagnetic fields as photons and  plasma effects.  Besides, one can employ the techniques handling source-induced GWs in literature to solve full second-order differential Eqs. \ref{eq:hplus}-\ref{eq:scalar} and find out the spectrum for $\omega<\kappa B$. Moreover,  dark energy effects could be considered  adopting suitable $f(R)$ models and electromagnetic fields. Finally,  a de-Sitter background   and the chameleon effect should  be assumed in this research line \cite{Burrage:2017qrf,Cembranos:2023ere,Tretyakov:2025cnk}.

\vspace{1cm}

{\bf Acknowledgments}. We thank Rome Samanta, Andrea Addazi, Gaetano Lambiase and Luca
Visinelli for useful discussions.  
We acknowledge the support of INFN sez. di Napoli, {\it iniziative specifiche} QGSKY and MOONLIGHT2. 
S.C. thanks the  {\it Gruppo Nazionale di Fisica Matematica} of {\it Istituto Nazionale di Alta Matematica} for the support.  This paper is based upon work from COST Action CA21136 - {\it Addressing observational tensions in cosmology with systematics and fundamental physics} (CosmoVerse), supported by COST (European Cooperation in Science and Technology).

\end{document}